\documentclass[useAMS,usenatbib,onecolumn,referee]{mn2e}

\usepackage{dcolumn}
\usepackage{graphicx}
\usepackage{url}
\usepackage{color}
\usepackage{pdflscape}

\newcommand{\cm}{cm$^{-1}$}

\newcommand{\X}{$X\,{}^1\Sigma^+$}
\newcommand{\A}{$A\,{}^1\Pi$}

\newcommand{\ai}{\textit{ab initio}}

\newcommand{\eqref}[1]{(\ref{#1})}

\newcommand{\Duo}{{\sc Duo}}
\newcommand{\Level}{{\sc Level}}
\newcommand{\xcross}{{\sc ExoCross}}
\newcommand{\PGOPHER}{{\sc PGOPHER}}

\title[ExoMol XXVIII: The spectrum of AlH]{ExoMol line lists XXVIII: The rovibronic spectrum of AlH}

\date{\today}
\author[Yurchenko et al]{\large Sergei N. Yurchenko, Henry Williams, Paul C. Leyland, Lorenzo Lodi, and Jonathan Tennyson\thanks{Email: j.tennyson@ucl.ac.uk} \\
Department of Physics and Astronomy, University College London, London WC1E 6BT, UK}
\date{Accepted XXXX. Received XXXX; in original form XXXX}

\pagerange{\pageref{firstpage}--\pageref{lastpage}} \pubyear{2018}

\begin{document}

\maketitle

\begin{abstract}


  A new line list for AlH is produced. The WYLLoT line list spans two
  electronic states \X\ and \A. A diabatic model is used to model the
  shallow potential energy curve of the \A\ state, which has a strong
  pre-dissociative character with only two bound vibrational states.
  Both potential energy curves are empirical and were obtained by
  fitting to experimentally derived energies of the \X\ and \A\
  electronic states using the diatomic nuclear motion codes \Level\
  and \Duo. High temperature line lists plus partition functions and lifetimes for three isotopologues $^{27}$AlH,  $^{27}$AlD and $^{26}$AlH
  were generated
  using \ai\ dipole moments. 
The line lists cover both the $X$--$X$ and $A$--$X$
  systems and are made available in electronic form at the CDS and
  ExoMol databases.

\end{abstract}

\label{firstpage}

\section{Introduction}

Aluminium is one of the commoner interstellar metallic  elements,
with a cosmic abundance of Al/H = $3 \times 10^{-6}$ but AlH has only been
rather sparingly observed. AlH was detected
in the photospheres of $\chi$ Cygni, a Mira-variable S-star, by \citet{56Herbig.AlH}
and much more recently around Mira-variable o Ceti by \citet{16KaWoSc.AlH}. AlH
was also detected in sunspots through lines in its \A\ -- \X\ electronic
band, which lies in the blue region of the visible \citep{00WaHiLi}; this spectrum was recently
analysed for its rotational temperature by \citet{10KaRaBa.AlH}.

AlH is difficult to detect in the interstellar medium because of its
small reduced mass, which causes its rotational transitions to occur
in the submillimetre region. This spectral region is typically
filtered by terrestrial atmospheric effects, therefore making it
difficult to detect from ground-based observations. So far searches
for interstellar AlH have proved negative with, for example, only an
upper limit set for the molecule rich IRC+10216 \citep{10CeDeBa.HCl}
using the Herschel Space Observatory.
\citet{10HaZixx.AlH,14HaZixx.AlH,16HaZixx.AlH} have undertaken
systematic improvement of the AlH submillimetre frequencies, including
hyperfine splittings, to aid future detections. The work of Halfen \&\ Ziurys is directly
complementary to the work presented here which is aimed at providing
ro-vibrational and rovibronic spectroscopic data. Other laboratory
studies of AlH spectra, both experimental and theoretical, are
discussed below.

Molecular spectra are useful for providing isotopic abundances.  Aluminium
 has only one stable isotope, $^{27}$Al, but $^{26}$Al has a long
half-life, in excess of 700~000 years.  The mass fraction ratio of
interstellar $^{27}$Al to $^{26}$Al can therefore  provide important
information on the formation of Al isotopes
\citep{84MaLiWh.AlH,08DiKnLi.AlH,12LuDoKa.AlH}; these could be probed
using the spectrum of AlH.  The AlH molecule is also thought to be an important
constituent of the atmospheres of so-called Lava-planets
\citep{jt693}.

The $A$--$X$ band has also been considered as a possible
means of producing ultra-cold AlH using laser cooling \citep{11WeLaxx.AlH}.
Lifetimes of the \A\ state were measured by \citet{79BaNexx.AlH}, while
\citet{03TaTaDa.AlH} considered lifetimes for the triplet system $b$--$a$.

\citet{88BaLaxx.AlH} used a full configuration interaction (FCI) \ai\ method to
produce \ai\ potential energy and  dipole moment curves of AlH, where they also 
estimated the dissociation 
energies of the \X\ and \A\ states and lifetimes of the two bound vibrational
sates of \A. The heat of formation of AlH was estimated by \citet{02Coxxxx.AlH}
using  DFT methods.
\citet{94CaJoAn.AlH} presented \ai\ estimates for the dipole moments and
transition dipole moments of AlH using a quasi-degenerate variational
perturbation
theory and averaged coupled-pair functional theory.

The ExoMol project \citep{jt528} aims to provide line lists of
spectroscopic transitions for key molecular species which are likely
to be important in the atmospheres of extrasolar planets and cool
stars.  \citet{13RaReAl.NaHAlH} analysed BT-Settl synthetic spectra
\citep{BT-Settl} for M-dwarf stars and suggested that the CaOH band at
5570 \AA, and AlH and NaH hydrides in the blue part of the spectra
constituted the main species still missing in the models.  An ExoMol
line list for NaH was subsequently computed by \citet{jt605}; here we
construct the corresponding line lists for isotopologues $^{27}$AlH,
$^{27}$AlD and $^{26}$AlH of aluminium hydride. 

We
previously provided line lists for isotopologues of AlO \citep{jt598};
this work follows closely on the methodology developed for treating
this open shell system \citep{jt589}. Here we consider transitions
within the \X--\X\ and electronic \A\ -- \X\ bands.  The
\A\ potential energy curve (PEC) is very shallow with a strong
pre-dissociative character and can accommodate only two vibrational
states  \citep{34HoHuxx.AlH} with a small barrier before the
dissociation. Here we apply the \Duo\ diatomic code \citep{Duo} to
solve the nuclear motion problem for the \X\ and \A\ coupled
electronic states of AlH and to generate a line list for the $X$--$X$
and $A$--$X$ bands using empirical PECs and high level \ai\
(transition) dipole moment curves (DMC). The centrifugal correction
due to the Born-Oppenheimer breakdown effect is also considered along
with an empirical electronic angular momentum coupling between \X\ and
\A.  The empirical PECs were obtained by fitting the corresponding
analytical representations to the experimental energies of AlH derived
from the measured line positions available in the literature using the
MARVEL (measured active rotation-vibration energy levels)  methodology \citep{jt412}. 
Special measures were taken to ensure that the unbound and quasi-bound states are not included in the line lists. Lifetimes  and partition functions are also provided as part of the line lists supplementary material, which are available from the CDS and ExoMol databases. Comparisons
with experimental spectra and lifetimes are presented.

The paper is structured as follows. Section~\ref{s:method} describes the methods used and includes a discussion of previous laboratory data, Section~\ref{s:results} presents our results and Section~\ref{s:concl} offers some conclusions.

\section{Method}
\label{s:method}

Rotation-vibration resolved lists for the ground \X\ and \A\ excited
electronic states of AlH were obtained by direct solution of the
nuclear-motion Schr\"{o}dinger equation using the \Duo\ program
\citep{Duo} in conjunction with  empirical
PECs and  \ai\ (transition) DMCs. In principle the calculations could be performed using \ai\
PECs and
coupling curves \citep{jt632}; however, in practice this does
not give accurate enough transition frequencies or wavefunctions so
the PECs was actually characterised by fitting to observed
spectroscopic data. Conversely, experience
\citep{jt573} suggests that retaining \ai\ diagonal and transition
dipole moment curves gives the best predicted transition intensities;
this approach is adopted here.

\subsection{Experimental data}

There has been considerable laboratory work on the spectrum of AlH and AlD.
High resolution studies considered for this work are summarised in
Table~\ref{tab:obsdata}. In addition there have been a recent studies involving higher 
electronic states of AlH by \citet{17SzHaKo.AlH} and \citet{17SzMoLa.AlH}.

\begin{landscape}

\begin{table}
\caption{Summary of high resolution spectroscopic data for AlH. FT = Fourier Transform.}
\label{tab:obsdata} \footnotesize
\begin{center}
\begin{tabular}{llllll}
\hline\hline
Reference&  Bands & Method     &  range in $J$/$v$&   Frequency range    & Comments \\
\hline
\citet{69Huronx.AlH}& \A\ -- \X\ & Absorption spectroscopy & $v = 0 - 1$ \\

\citet{78RaBaKh.AlH}& b~$^3\Sigma^+$ -- a~$^3\Pi$ & Emission spectroscopy & $J = 3 - 16$&
1-1 band at 3808 \AA &  3 P-branches, 3 R-branches, unresolved Q\\

\citet{79BaNexx.AlH} &\A\ -- \X\ &Dye laser excitation& $v = 0 - 1$,$J = 6 - 26$\\

\citet{87DeNeRa.AlH} & $\Delta v = 2$  & Furnace emission &  $v = 0 - 8$, $J = 1 - 34$&  \\

\citet{88ZhStxx.AlH}& C~$^1\Sigma^+$ -- \X\ &Dye laser spectrocopy&  $v = 0 - 1$\\
                    & D~$^1\Sigma^+$ -- \X\\

\citet{92UrJoxx.AlH}& vibrations&  Infrared diode laser&  $v = 0 - 7$ && AlD\\

\citet{92RiPaNe.AlH}&\A\ -- \X\ & Emission spectra & $v=0,1$ & 13 -- 30 nm &
0-0, 0-1, 1-0, 1-1, 1-2 bands\\

\citet{92YaHixx.AlH}&vibrations&  Infrared diode laser&  $v = 0 - 7$, $J=2 - 31$ &&
AlH \&\ AlD; 1-0, 2-1,
3-2, 4-3\\

\citet{92ZhShGr.AlH}& C~$^1\Sigma^+$ -- \X\ &laser spectrocopy&  $v = 0 - 1$, $J = 0 - 18$\\
                    & b~$^3\Sigma^-$ -- \X\\

\citet{93WhDuBe.AlH} &$\Delta v = 1$& Infrared emission &$v = 0 - 7$\\

\citet{94ItNaTa.AlH} &$\Delta v = 1$& Infrared emission &$v = 0 - 3$, $J=0-22$& 1400 -- 1800 \cm \\

\citet{95GoSaxx.AlH} &rotations &Submillimeter & $J=0-1$& 387 GHz& hyperfine resolved\\

\citet{96RaBexx.AlH} &\A\ -- \X\ &FT emission &  $v = 0 - 1$, $J = 0 - 18$\\

\citet{98YaDaxx.AlH} &\A\ -- \X\ &Laser fluorescence & $v = 0 - 4$&\\

\citet{00NiDaxx.AlH}&\A\ -- \X\ &Fluorescence emission & $v=0$, $J=4-14$&&
e \&\ f doublets resolved.\\

\citet{03TaTaDa.AlH} &b~$^3\Sigma^-$ -- a~$^3\Pi$ &Laser fluorescence & $v = 0 - 1$, $J=0-6$&  26222 -- 26400 \cm &
$J=1$ lifetime measured\\

\citet{04HaZixx.AlH}&rotations &Microwave FT & $J=0-2$& 377--393 GHz& hyperfine resolved\\

\citet{09SzZaxx.AlH} &\A\ -- \X\ && $v=0-3$& 18 000 -- 25 000 \cm &\A\ $v = 1$, $J = 5$
state perturbed\\

\citet{10HaZixx.AlH} &rotations &Direct absorption & $J=0-4$& 393-590 GHz& AlD\\

\citet{10SzZaxx.AlH}& C~$^1\Sigma^+$ -- \X\ && $v = 0 - 5$, $J= 0-28$& 42000 -- 45000 \cm\\

\citet{11SzZaxx.AlH}& C~$^1\Sigma^+$ -- \A\ && $v = 0 - 3$, $J= 0-18$ & 20000 -- 21500 \cm \\
                    &\A\ -- \X\\

\citet{14HaZixx.AlH}& rotations &Direct absorption & $J=1-4$& 755 -- 787 GHz & AlH \&\ AlD\\

\citet{15SzZaHab.AlH}&\A\ -- \X& Optical dispersion& $v=0,1$, $J=0-29$&22400 -- 23,700 \cm & AlD\\
\citet{17SzMoLa.AlH}&\A\ -- \X,C~$^1\Sigma^+$ -- \A\ & FT emission& $v=0,1$, $J=0-29$&22400 -- 23,700 \cm & AlD\\
\hline
\end{tabular}
\end{center}
\end{table}
\end{landscape}

Transition frequencies of AlH were collected from papers by
\citet{87DeNeRa.AlH,92YaHixx.AlH,93WhDuBe.AlH,94ItNaTa.AlH,96RaBexx.AlH,09SzZaxx.AlH,04HaZixx.AlH,11SzZaHa.AlH,14HaZixx.AlH}
listed in Table~\ref{tab:obsdata} which concerned $^{27}$AlH and $^{27}$AlD and transitions within the ground electronic state or the \A\ -- \X\ band.
The transitions were used as input for a MARVEL analysis \citep{jt412,12FuCsa}.
Much of the data
\citep{92YaHixx.AlH,94ItNaTa.AlH,96RaBexx.AlH,09SzZaxx.AlH,11SzZaHa.AlH}
was validated by MARVEL requiring, at most,  small
uadjustments of the assigned uncertainties to make them consistent with each
other.

The work of \citet{04HaZixx.AlH,14HaZixx.AlH} is hyperfine-resolved but hyperfine
splittings are not present in the other studies and are not considered in this work.

The works of \citet{93WhDuBe.AlH} and \citet{87DeNeRa.AlH} are important as the source of high $v$ numbers (up to $v=5$ and $v=8$, respectively) in the \X\ state. 


Running MARVEL on this network of 917
validated transitions gave 331  empirical energy levels, 283 in the \X\ state and 48 in the \A\ state of $^{27}$AlH . For the
\X\ state, $J$ spanned the range 0 to 40 and $v$ went from 0 to 8. For
the \A\ state, $J$ spanned the range 0 to 29 and $v$ only included 0 and 1. Note that the \A\ state state is very shallow and supports, at most, only these two vibrational states. This issue is discussed further below.

In the case of  $^{27}$AlD, 581 experimental transition frequencies  $X$--$X$ were taken from \cite{92UrJoxx.AlH,92YaHixx.AlH,04HaZixx.AlH} and used in MARVEL to produce 301 term values of \X\ covering $v=0$--7 and $J=0$\ldots 39. The $A$ state term values were taken directly from \citet{15SzZaHab.AlH} ($v=0,1$, $J_{\rm max} = 29$).


\subsection{Potential energy and dipole moment curves}

\begin{figure}
\begin{center}
\includegraphics[width=0.48\textwidth]{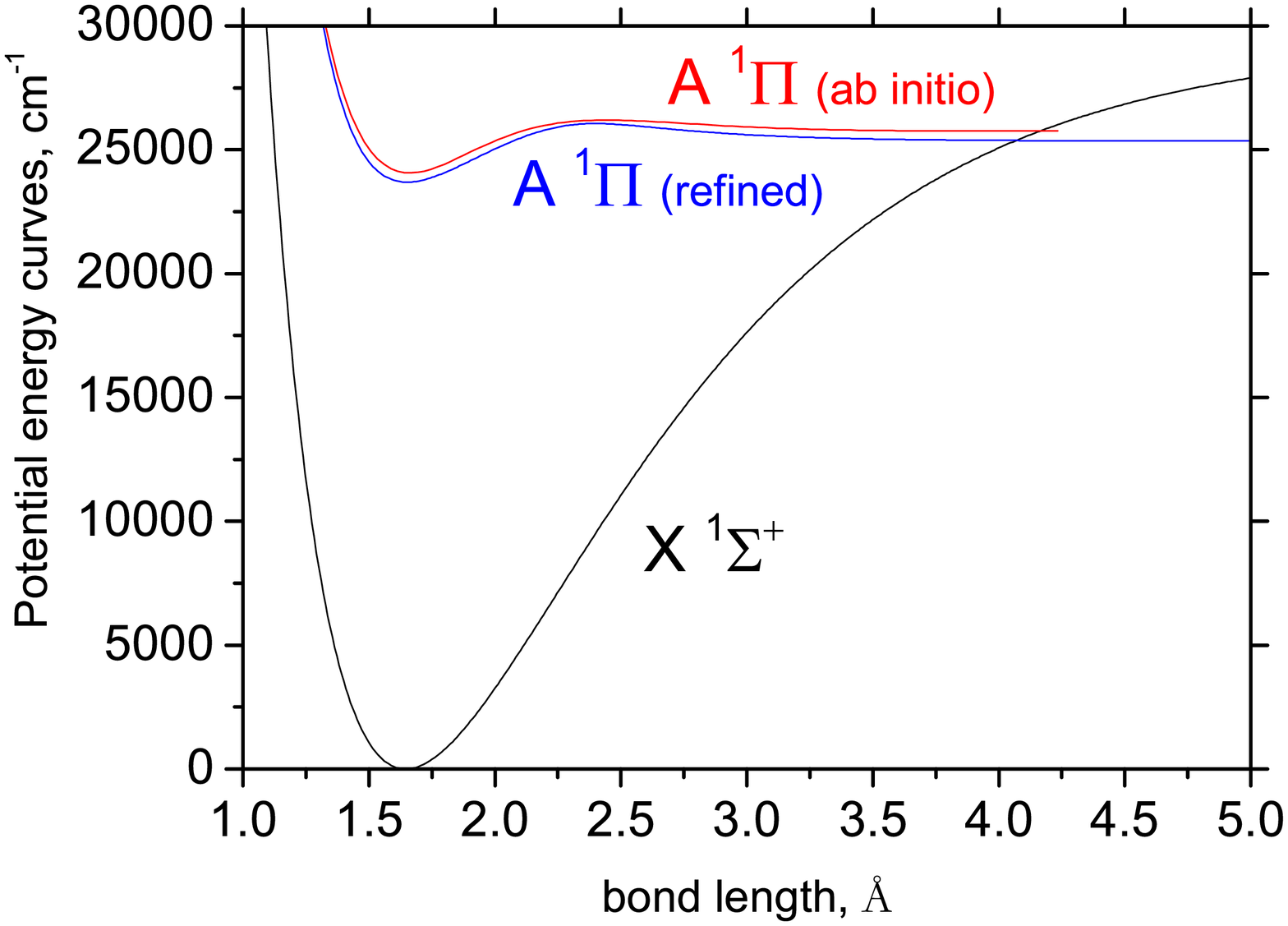}
\includegraphics[width=0.48\textwidth]{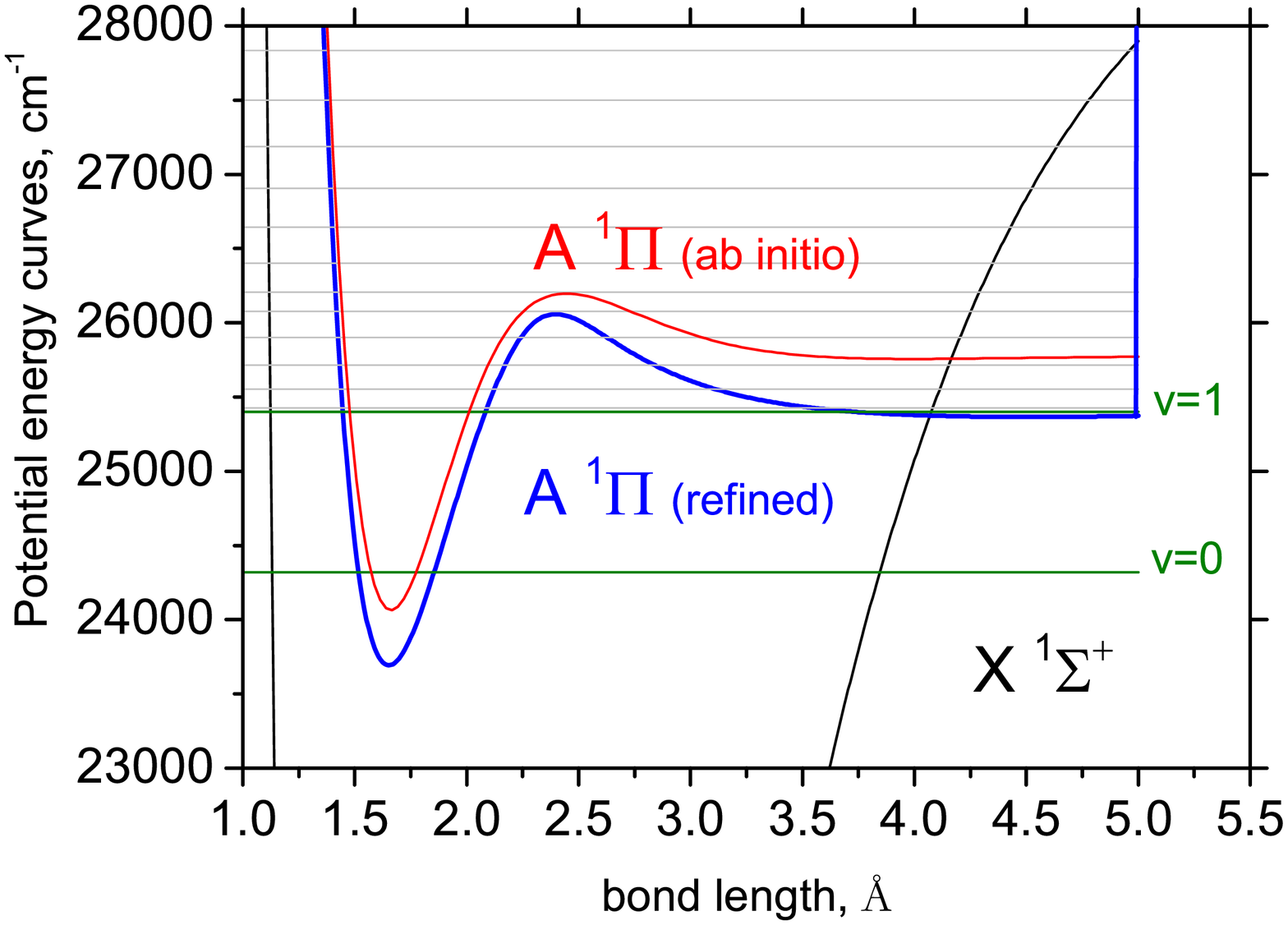}
\caption{AlH \X\ and \A\ state potential energy curves. The zoomed
figure (left) also shows the AlH vibrational states and \Duo\
vibrational basis functions, see text for details.}
\label{fig:PECs}
\end{center}
\end{figure}

There are a number of previous studies of the \A\ and \X\ curves of
AlH, see \citet{13BrWaxx.AlH}, \citet{14SeHoLi.AlH}, and references
therein. The ground electronic PEC has a nice Morse-like structure.
Experimental data on the $X$ state cover 
vibrational excitations up to $v=8$; therefore we decided to obtain
the $X$-state PEC fully empirically by fitting it  to the experimental
frequencies from \cite{87DeNeRa.AlH,93WhDuBe.AlH,94ItNaTa.AlH}. The $X$-state PEC was
represented using the Extended Morse Oscillator (EMO) potential
\citep{EMO} given by
\begin{equation}
\label{eq:EMO}
V(r) = V_{\rm e} + (A_{\rm e}-V_{\rm e})\Bigg[1-\exp\Bigg(-\beta_{\rm EMO}(r)(r-r_{\rm e})\Bigg)\Bigg]^2 ,
\end{equation}
where $(A_{\rm e}-V_{\rm e})$ is the dissociation energy, $V_{\rm e}$ is the minimum of the PEC, which for the \X\ state was set to zero,  $N$ is the expansion
order parameter,
$r_{\rm e}$ is the equilibrium internuclear bond distance,
$\beta_{\rm EMO}$ is the distance dependent exponent coefficient,
defined as
\begin{equation}
\label{eq:beta}
\beta_{\rm EMO}(r) = \sum\limits_{i=0}^{N} B_i \xi_{p} (r)^i
\end{equation}
and ${\xi_{p}}$ is the \u{S}urkus variable \citep{84SuRaBo.method} given by
\begin{equation}
\label{eq:y}
\xi_{p}(r) =  \frac{r^p - r^{p}_{\rm ref}}{r^p + r^{p}_{\rm ref}}
\end{equation}
with $p$ as a parameter.  Use of the EMO has two advantages. First, it guarantees a
correct dissociation limit and second, allows extra flexibility in the degree
of the polynomial around a reference position $r_{\rm ref}$, which was defined as the
equilibrium internuclear separation ($r_{\rm e}$) in this case.
Figure~\ref{fig:PECs} shows our empirical PEC of AlH in its $X$ state.

This closed shell \X\ ground state was fitted to an EMO using Level \citep{level}. To allow for rotational Born-Oppenheimer breakdown (BOB) effects \citep{lr07} which become important
for $J > 20$, the vibrational kinetic energy operator was extended by
\begin{equation}
\label{e:dist}
-\frac{\hbar^2}{2 \mu r^2} \to -\frac{\hbar^2}{2 \mu r^2} \left( 1 + g^{\rm BOB}(r)\right),
\end{equation}
where the unitless BOB functions $g^{\rm BOB}$ are represented by the polynomial
\begin{equation}
 g^{\rm BOB}(r)= \left[ (1 - \xi_p) \sum_{k\ge 0}^{N_{T}} A_k \xi_p^k  + \xi_p A_{\rm \infty} \right]
\label{eq:bobleroy}
\end{equation}
where $\xi_p$ as the \u{S}urkus variable and $p$, $A_k$ and  $A_{\rm \infty}$ are adjustable parameters.

\begin{figure}
\begin{center}
\includegraphics[width=0.75\textwidth]{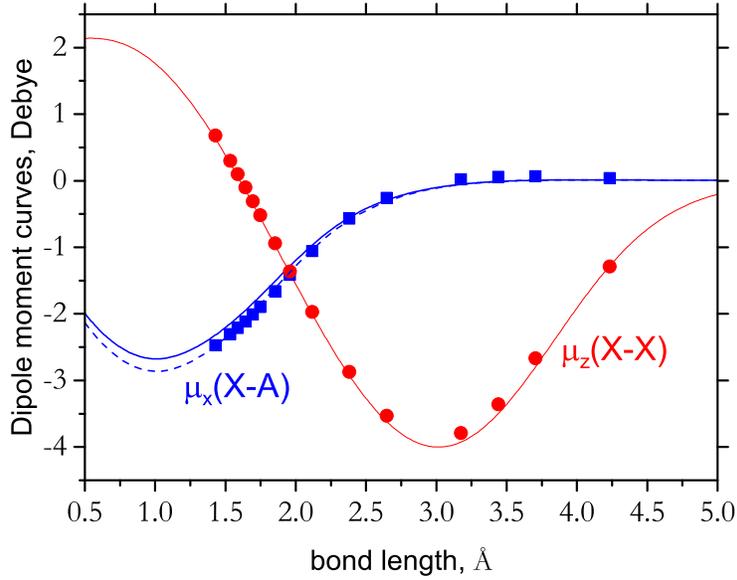}
\caption{ {\it Ab initio} (transition) dipole moment curves for AlH,  $X$--$X$ and
$A$ -- $X$: The solid lines are from this work and the symbols represent the dipoles of \citet{88BaLaxx.AlH}; the dashed curve is a scaled version of
our \ai\ transtion dipole, see text for details. }
\label{fig:dipole}
\end{center}
\end{figure}

Given the shallow nature of the $A$ curve which also appears to
undergo an avoided crossing we decided to perform our own calculations
using a high level of electronic structure theory. {\it Ab initio}
PECs and Dipole Moment Curves (DMCs) were computed using the MOLPRO
electronic structure package \citep{MOLPRO} at the multi-reference
configuration interaction (MRCI) level using an aug-cc-pV5Z Gaussian
basis set. Calculations were performed at 120 bond lengths over the
range of $r$ = 2 to 8 a$_0$.  Figure~\ref{fig:PECs} shows the \ai\ PEC
of \A, which only supports two bound vibrational states. It also shows
a maximum at about 4.5 $a_0$ which is probably associated with an
avoided crossing. Figure~\ref{fig:dipole} shows our \ai\ DMCs, which
agree well with the \ai\ dipole moment values from
\citet{88BaLaxx.AlH}, although, as discussed below, the magnitude of
our transition dipole is slightly smaller.  Our calculations give a
permanent dipole moment of 0.158 D (absolute value) at $r = 1.646$
\AA\ at equilibrium, which is slightly higher than that by
\citet{88BaLaxx.AlH}, 0.12~D. This is significantly less than the
absolute value of 0.186 used by CDMS \citep{CDMS}, which is taken from
an old calculation by \citet{75MeRoxx.AlH}. We also note that
the $X$ dipole also changes sign close to equilibrium. We return to
these issues below. \citet{88MaRoSa.AlH} in their \ai\ work showed a
strong variation of the dipole and obtain a value of 0.3~D for $\mu_0$
(i.e. a vibrational averaged in the ground vibrational state), while
our value is 0.248~D. No experimental values exist.  The \A\ -- \X\
transition dipole moment of AlH also undergoes a change in behaviour
in the region around 4.5 $a_0$.

In order to represent the complex shape of the shallow \A\ potential energy curve (see Fig.~\ref{fig:PECs}), we used a diabatic-like scheme, where the effect of the avoided crossing is described by a $2\times 2$ matrix:
\begin{equation}\label{e:V1W/WV2}
\underline{B} = \left(
\begin{array}{cc}
  V_1(r) & W(r) \\
  W(r) & V_2(r)
\end{array}
\right).
\end{equation}
Here $V_1(r)$ is given by the EMO potential function in Eq.~\eqref{eq:EMO}, while $V_2(r)$ is represented by a simple repulsive form
$$
V_2(r) = \frac{w_{6}}{r^6}.
$$
The coupling $W(r)$ is given by
\begin{equation}\label{e:W(r)}
W(r) = W_0 + \frac{\sum_{i\ge 0} w_{i} (r-r_{\rm cr})^i  }{\cosh[\beta (r-r_{\rm cr})]},
\end{equation}
where $r_{\rm cr}$ is a crossing point. The two eigenvalues  of $\underline{B}$ are given by
\begin{eqnarray}
  V_{\rm low}(r) &=& \frac{V_1(r)+V_2(r)}{2}-\frac{\sqrt{[V_1(r)-V_2(r)]^2+4 \, W^2(r)}}{2}, \\
  V_{\rm upp}(r) &=& \frac{V_1(r)+V_2(r)}{2}+\frac{\sqrt{[V_1(r)-V_2(r)]^2+4 \, W^2(r)}}{2}, \\
\end{eqnarray}
where the lowest root $V_{\rm low}(r)$ corresponds to the \A\ adiabatic PEC.

Initially, the expansion parameters representing this form were
obtained by fitting to the \ai\ $A$-state PEC shown in
Fig.~\ref{fig:PECs} and then refined by fitting to the MARVEL
energies. Currently \Duo\ does not support quasi-bound or continuum
solutions, see \citet{Duo}. Technically, by virtue of the sinc DVR (discrete variable representation)
method used by \Duo\ to solve the vibrational ($J=0$) Schr\"{o}dinger
equation, \Duo\ uses infinite walls at each end of the integration
grid (0.5 -- 5.0~\AA\ in our case) as boundary conditions, which is
illustrated in Fig.~\ref{fig:PECs}.  For the \A\ state, we selected 20
basis functions generated by solving the $J=0$ problem for the PEC of
the \A\ state using the sinc DVR method (and 60 vibration basis
functions for \X). All these basis set functions have zeros at the boundaries, $r=$0.5 and 5.0~\AA,
and thus represent bound wavefunctions.
Therefore the solution of the coupled rovibronic Schr\"{o}dinger
equations with this basis set contains a mixture of real \A\ bound
states ($v=0$ and $v=1$) and a large number of 
continuum states. The corresponding energies of these states for $J=0$ are shown in Fig.~\ref{fig:PECs} as horizontal
lines. From these solutions only the $v=0$ and $v=1$ states are actually bound and thus selected for the final line list. The bound states can be easily
distinguished from the quasi- or unbound states using
the $A$--$X$ transition probabilities or lifetimes, which differ by
3--4 orders of magnitude.  It should be noted that the \A\ computed
energies suffer from accidental resonances at higher values of $J$
between bound and unbound states, which affect the accuracy of the
calculated $v=0$ and $v=1$ energies. As shown below, this
is partly resolved by replacing the theoretical energies with the
experimentally derived (MARVEL) values.

\begin{figure}
\begin{center}
\includegraphics[width=0.75\textwidth]{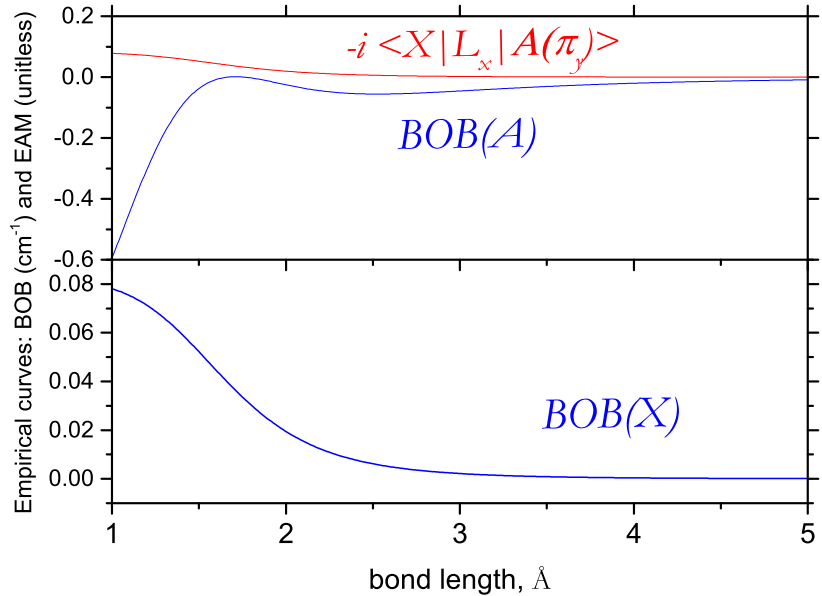}
\caption{Empirical curves for AlH: Born-Openheimer Breakdown (BOB) for the
\X\ and \A\ states, plus the  $L_x$ electronic angular momentum coupling
between the two states.}
\label{fig:BOB}
\end{center}
\end{figure}

The BOB-correction in the form given in Eq.~\eqref{eq:bobleroy} was used for the $A$-state as well.
In these fits the $X$-state parameters were fixed to the values obtained using Level. The BOB-curves are shown in Fig.~\ref{fig:BOB}.

In order to account for the $\Lambda$-doubling effect, we also used an
empirical electronic angular momentum (EAM) coupling between the \A\
and \X\ states, which was represented by
\begin{equation}
 H^{\rm EAM}(R)= (1 - \xi)  A_0^{\rm EAM},
\label{eq:EAM}
\end{equation}
which is nothing else than Eq.~\eqref{eq:bobleroy} truncated after the leading term.
The final value of $A_0^{\rm EAM} $ is 0.1475~\cm\ and the EAM curve is shown in Fig.~\ref{fig:BOB}.

The final fit gave an observed minus calculated  root-mean-square (rms) error
of 0.025  \cm, when compared to our MARVEL energy levels for the \X\ state. The MARVEL energy levels of the \A\ state are reproduced with an rms error of 0.59~\cm.


\citet{79BaNexx.AlH} reported AlH dissociation energies measured using a
hollow cathode discharge by dye laser excitation, 3.16 $\pm
0.01$~eV and 0.24 $\pm 0.01$~eV for the \X\ and \A\ states, respectively.
The dissociation energy ($D_{\rm e}$) of our empirical PEC of the \X\
state is 3.644~eV which overestimates the experimental value. This should not be a problem for our line list since
the contributions from the highly excited vibrational states of \X\  is
practically zero at such energies.  For the \A\ PEC we obtained   $D_{\rm e}$ = 0.209~eV (\ai) and 0.210~eV (refined PEC), which compare
well to the experimental value by \citet{79BaNexx.AlH}.
It should be noted that
\citet{88BaLaxx.AlH} also reported  \ai\ FCI dissociation energies
which coincide with the experimental values by \citet{79BaNexx.AlH}.

Since the current version of \Duo\ does not account for the isotopic-effect explicitly and thus is not capable of treating a mass-independent model as, for example, in \Level, we had to create independent models for different isotopologues. Therefore the same fitting procedure was repeated for AlD, where the model curves were fitted to the AlD experimentally derived energies (MARVEL). Fortunately, the experimental data set for AlD is almost as large as that for AlH.

Final parameters for all the curves representing our two spectroscopic models for AlH and AlD (PECs, DMCs, and other empirical curves) and used in \Duo\ are given in the supplementary material in the form of the \Duo\ input. The program \Duo\ is freely available via the {www.exomol.com} web site. The actual curves can be extracted from the \Duo\ outputs, which are also provided.


\subsection{Lifetimes}

The lifetimes of AlH in the \A\ state were measured by
\citet{79BaNexx.AlH} using a hollow cathode discharge by dye laser
excitation, who reported two values: 66$\pm 4 $ ns ($v=0,J=7$) and
83$\pm 6$ ($v=1,J=5$). Using our Einstein coefficients and program
\xcross\ \citep{jt708} we obtained 73.64 ~ns and 102.53~ns for these
states, respectively.  These lifetimes can also be compared to
theoretical values given by \citet{88BaLaxx.AlH} of 64.3 ns and 96.6
ns, respectively. Our lifetimes are about 1.12 times higher than
experiment, which indicates that our transition dipole moment $A$--$X$
is about 1.07 times too strong. Considering that the $A$--$X$ curve by
\citet{88BaLaxx.AlH} is also the same factor larger, see in
Fig.~\ref{fig:dipole}, we have decided to scale our \ai\ transition
dipole moment $A$--$X$ by 1.07 up. The resulted transition dipole
moment is shown in Fig.~\ref{fig:dipole}, where it matches better the
\ai\ TDMC by \citet{88BaLaxx.AlH}. Using the scaled TDMC, we also
obtain a better match for the lifetimes: 64.3~ns ($v=0,J=7$) and 89.6
($v=1,J=5$). This transition DMC is put forward to produce the AlH
line lists.

\begin{figure}
\begin{center}
\scalebox{0.6}{\includegraphics{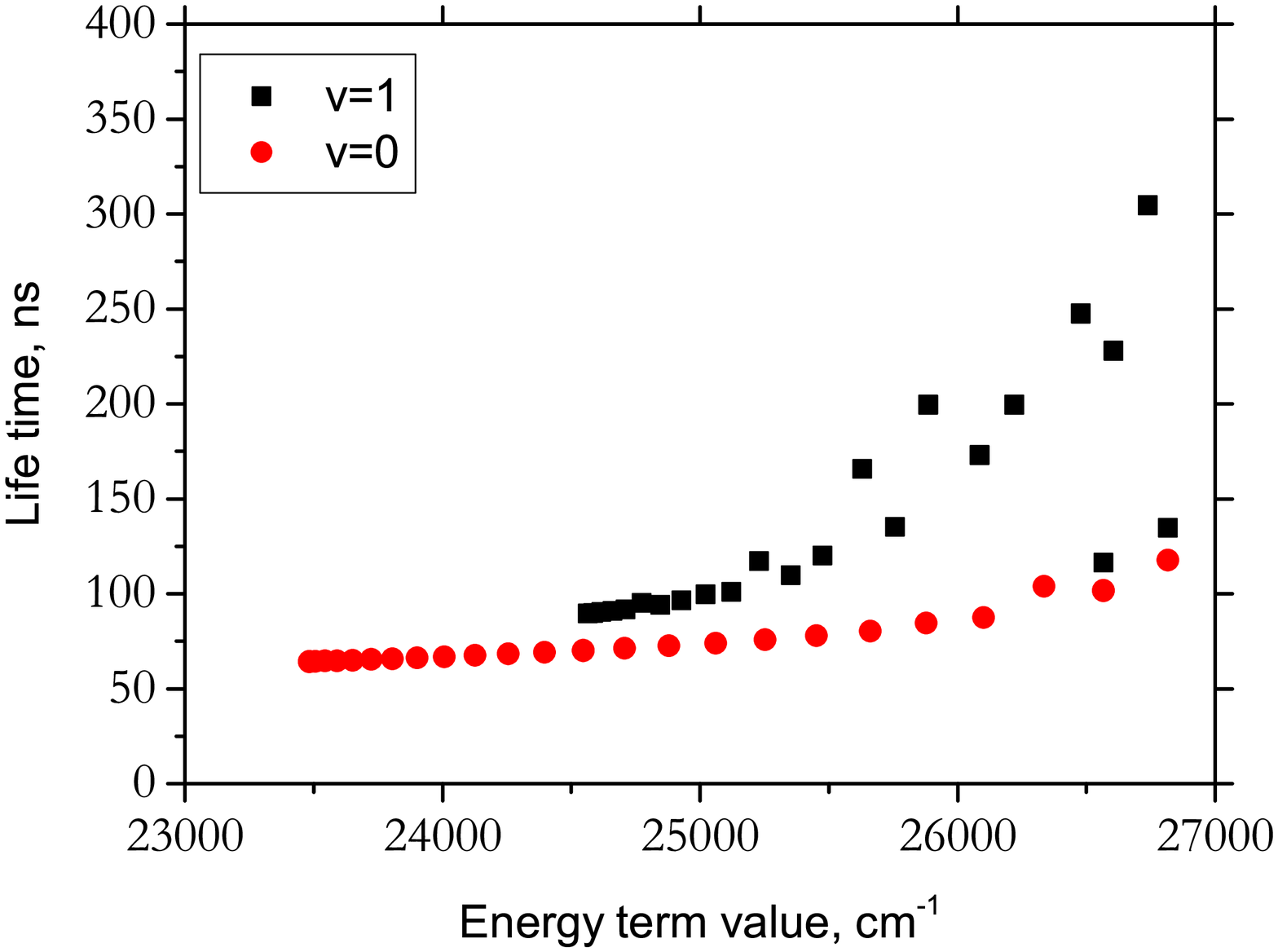}}
\caption{Lifetimes of AlH in the \A\ state. }
\label{f:life}
\end{center}
\end{figure}

 Figure \ref{f:life} shows our lifetimes of the $v=0$ and
$v=1$ ($A$) rovibronic states of AlH using the scaled DMC. The oscillations in the $v=1$ progression is probably due to accidental resonances 
with  unbound
states (see below).


\subsection{Line list generation}

Line lists for AlH and AlD were generated using the program \Duo.

In order to reduce the numerical noise in the intensity calculations of high overtones characterized 
by small transition probabilities  in the spectra of the \X\ state (see recent recommendations by \citet{16MeMeSt}) the DMCs are represented analytically. We use the following expansion \citep{jt703,jt711}:
\begin{equation}
\label{e:mu}
\mu(r)=\sum^{N}_{k=0}\mu_{k}\, z^{k} (1-\xi_p),
\end{equation}
where $z$ is  the
damped-coordinate given by:
\begin{equation}\label{e:damp}
z = (r-r_{\rm ref})\, e^{-\beta_2 (r-r_{\rm ref})^2-\beta_4 (r - r_{\rm ref})^4}.
\end{equation}
Here $r_{\rm ref}$ is a reference position equal to $r_{\rm e}$ by default and $\beta_2$ and $\beta_4$ are damping factors. The expansion parameters are given in the supplementary material. 
As an additional measure to reduce numerical noise in the overtone intensities,
a dipole moment cutoff of $10^{-7}$~D was applied to the vibrational dipole moments: 
all transitions for which the vibrational dipole moments are smaller than $10^{-7}$~D  were ignored.

The \A\ -- \X\ (bound) spectrum only contains transitions to/from the upper states $v'=0$ and $v'=1$, i.e. no  overtones, 
and thus should not suffer from the numerical noise issue as much as the \X\ -- \X\ band.
Therefore the \A--\X\ transition dipole
moment was given directly in the (scaled) \ai\ grid representation of 120 points. The latter points are interpolated by \Duo\ onto the sinc DVR  grid using the cubic splines method (see \citet{Duo}
for details).

Only bound vibrational and rotational states were retained which meant for $^{27}$AlH considering $J \leq 82$ for the \X\ state and $J \leq 25$ for the \A\ state. For $^{27}$AlD the range of $J$ was increased to $J_{\rm max}= $ 108 and $35$, respectively. \Duo\ input files used to generate the line lists are included as part of the supplementary data. This procedure was then
simply repeated for $^{26}$AlH by changing the mass of Al and nuclear statistics factor from 12 to 22. 
All \A\  empirical energies in  the .states file of $^{27}$AlH were replaced by the MARVEL values, or by values generated using \PGOPHER\ from the constants by \citet{15SzZaHab.AlH} if the MARVEL energies were not available; for
$^{27}$AlD we used the experimentally derived term values by \citet{15SzZaHab.AlH}.

The line lists  are stored in the standard ExoMol format \citep{jt631} which involves a states file listing
all the levels and a transitions file giving the Einstein A coefficients for each transition. The $^{27}$AlH line list contains 1,551 states and 39,483 transitions;
the $^{27}$AlD line list contains 2,930 states and 85,494 transitions; The $^{26}$AlH line list contains 1,549 states and 35,910 transitions.

 Tables~\ref{tab:levels}
and \ref{tab:trans} give samples of these files. Full versions can be found at
 CDS, via
ftp://cdsarc.u-strasbg.fr/pub/cats/J/MNRAS/, or
http://cdsarc.u-strasbg.fr/viz-bin/qcat?J/MNRAS/, or from
www.exomol.com.


\begin{table*}
\caption{Extract from the state file for $^{27}$Al$^{1}$H. Full tables
are available from http://cdsarc.u-strasbg.fr/cgi-bin/VizieR?-source=J/MNRAS/xxx/yy and www.exomol.com.}

\label{tab:levels}
\tt
{\tt  \begin{tabular}{rrrrrcclrrrr} \hline \hline
$n$ & Energy (\cm) & $g_i$ & $J$ & $\tau$ &  Parity & e/f	& State	& $v$	&${\Lambda}$ &	${\Sigma}$ & $\Omega$ \\ \hline
 1&     0.000000 &    12    &   0 &    inf      &  +   &   e &  X1Sigma+   &   0 &   0  &  0  &  0 \\
 2&  1625.069321 &    12    &   0 &  4.9283E-03 &  +   &   e &  X1Sigma+   &   1 &   0  &  0  &  0 \\
 3&  3194.213685 &    12    &   0 &  2.6224E-03 &  +   &   e &  X1Sigma+   &   2 &   0  &  0  &  0 \\
 4&  4708.817022 &    12    &   0 &  1.8630E-03 &  +   &   e &  X1Sigma+   &   3 &   0  &  0  &  0 \\
 5&  6170.193041 &    12    &   0 &  1.4908E-03 &  +   &   e &  X1Sigma+   &   4 &   0  &  0  &  0 \\
 6&  7579.564189 &    12    &   0 &  1.2739E-03 &  +   &   e &  X1Sigma+   &   5 &   0  &  0  &  0 \\
 7&  8938.046805 &    12    &   0 &  1.1350E-03 &  +   &   e &  X1Sigma+   &   6 &   0  &  0  &  0 \\
 8& 10246.644027 &    12    &   0 &  1.0410E-03 &  +   &   e &  X1Sigma+   &   7 &   0  &  0  &  0 \\
 9& 11506.245734 &    12    &   0 &  9.7519E-04 &  +   &   e &  X1Sigma+   &   8 &   0  &  0  &  0 \\
10& 12717.633797 &    12    &   0 &  9.2834E-04 &  +   &   e &  X1Sigma+   &   9 &   0  &  0  &  0 \\
\hline
\hline
\end{tabular}}
\mbox{}\\

{\flushleft
$n$:   State counting number.     \\
$\tilde{E}$: State energy in \cm. \\
$g_i$:  Total statistical weight, equal to ${g_{\rm ns}(2J + 1)}$.     \\
$J$: Total angular momentum.\\
$\tau$: Lifetime (s$^{-1}$).\\
$+/-$:   Total parity. \\
$e/f$:   Rotationless parity. \\
State: Electronic state.\\
$v$:   State vibrational quantum number. \\
$\Lambda$:  Projection of the electronic angular momentum. \\
$\Sigma$:   Projection of the electronic spin. \\
$\Omega$:   Projection of the total angular momentum, $\Omega=\Lambda+\Sigma$. \\
}
\end{table*}

\begin{table}
\caption{Extracts from the transitions file for $^{27}$AlH.
Full tables
are available from http://cdsarc.u-strasbg.fr/cgi-bin/VizieR?-source=J/MNRAS/xxx/yy and www.exomol.co. }
\label{tab:trans}
\begin{center}
\begin{tabular}{rrrr}
\hline
       \multicolumn{1}{c}{$F$}  &  \multicolumn{1}{c}{$I$} & \multicolumn{1}{c}{$A_{FI}$ }& \multicolumn{1}{c}{$\tilde{\nu}_{FI}$} \\
\hline
        1075  &      1087  &  0.5502E-09   &          47.949528 \\
        1529  &      1488  &  1.5634E-11   &          47.965895 \\
         561  &       520  &  6.1183E-02   &          47.984746 \\
         218  &       176  &  0.6544E-09   &          48.007351 \\
        1415  &      1384  &  1.4996E-11   &          48.171601 \\
         441  &       399  &  1.1015E-01   &          48.615068 \\
         198  &       156  &  0.8765E-09   &          48.846517 \\
         198  &       156  &  3.0699E-03   &          48.846517 \\
\hline
\end{tabular}
\noindent
\mbox{}\\
{\flushleft
 $F$: Upper state counting number;\\
$I$:      Lower state counting number;\\
$A_{FI}$:  Einstein A coefficient in s$^{-1}$.\\
$\tilde{\nu}_{FI}$:  Energy term value in \cm.\\
}
\end{center}
\end{table}

\section{Results}
\label{s:results}

\subsection{Partition function}

\begin{figure}
\centering
\includegraphics[width=320pt]{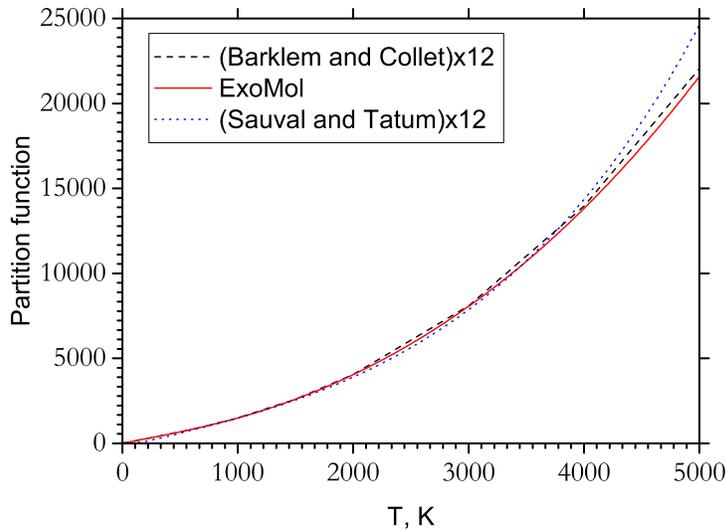}
\caption{Temperature dependence of the partition function of AlH computed using our line lists and compared to those by \protect\citet{84SaTaxx.partfunc} and \protect\citet{16BaCoxx.partfunc}.}
\label{f:pf}
\end{figure}

Partition functions were generated for each isotopologue by explicit summation of the energy
levels. Comparison for $^{27}$AlH with the recent results of \citet{16BaCoxx.partfunc}  and with the partition function generated using parameters from \citet{84SaTaxx.partfunc} shows
excellent agreement for temperatures below 5000~K (see Fig.~\ref{f:pf}) once allowance is made for the fact that ExoMol adopts the HITRAN convention \citep{jt692} which includes the full nuclear spin degeneracy
factor  in the partition function (12 in case of $^{27}$AlH, 18  in case of  $^{27}$AlD and 22 in case of $^{26}$AlH).

It should be noted that AlH is unlikely to
be important at temperatures above 5000~K.

Partition functions, $Q(T)$,  on a 1 K grid up to 5000 K are given for each isotopologue in the
supplementary material. For ease of use we also provide
fits in the form proposed by \citet{jt263}:
\begin{equation}
\log_{10} Q(T) = \sum_{n=0}^6 a_n \left[\log T\right]^n \label{eq:pffit}
\end{equation}
with the values given in Table~\ref{tab:pffit}.

\begin{table}
\caption{Fitting parameters used to represent the partition functions,
see eq.~(\ref{eq:pffit}). Fits are valid for temperatures up to 5000~K.
}
\label{tab:pffit}
\begin{center}
\begin{tabular}{lrrr}
\hline
\hline
Parameter & \multicolumn{1}{c}{$^{27}$AlH} &  \multicolumn{1}{c}{ $^{27}$AlD} &  \multicolumn{1}{c}{ $^{26}$AlH}  \\
\hline
$       a_0     $&$      1.07862304206   $&$         1.25534381874   $&$         1.34186025327       $\\
$       a_1     $&$      0.38191971276   $&$         0.22993489470   $&$         0.38228272653       $\\
$       a_2     $&$     -2.22562466705   $&$        -1.92470848538   $&$        -2.22605038250       $\\
$       a_3     $&$      4.34557816661   $&$         6.16172026680   $&$         4.34313450679       $\\
$       a_4     $&$     -3.62598781682   $&$        -7.66641279825   $&$        -3.62196458297       $\\
$       a_5     $&$      1.71637190159   $&$         5.35122144723   $&$         1.71374588820       $\\
$       a_6     $&$     -0.47932820561   $&$        -2.24207082067   $&$        -0.47845359880       $\\
$       a_7     $&$      0.07359563236   $&$         0.55715405704   $&$         0.07344763428       $\\
$       a_8     $&$     -0.00476226937   $&$        -0.07543604728   $&$        -0.00475215090       $\\
$       a_9     $&$                      $&$         0.00427936817   $&$                             $\\
\hline
\end{tabular}
\end{center}
\end{table}

\subsection{Spectra}

\begin{figure}
\begin{center}
\includegraphics[width=0.9\textwidth]{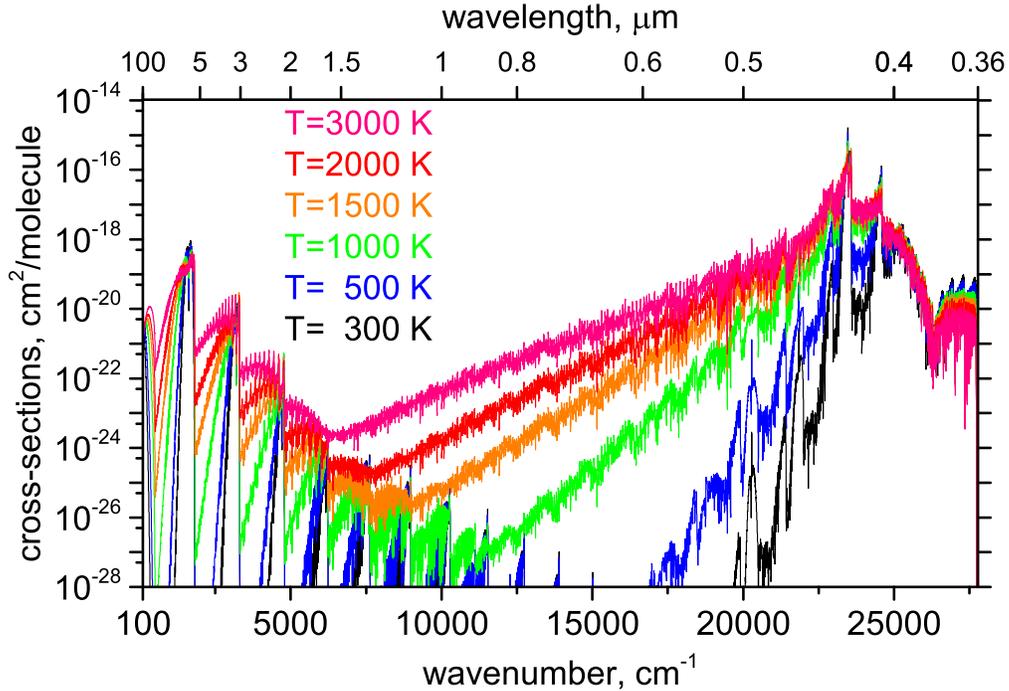}
\caption{Temperature dependence of the AlH absorption spectrum using the Gaussian profile with HWHM=5~\cm. The absorption profile becomes flatter with
increasing temperature.}
\label{f:Temp}
\end{center}
\end{figure}

In the following, we present different spectra of AlH computed using the new line lists and utilizing the  program \xcross\ \citep{jt708}.
Figure~\ref{f:Temp} gives an overview of the AlH line list in the form of absorption cross sections for a range of temperatures from 300 to 3000~K.

\begin{figure}
\begin{center}
\includegraphics[width=0.5\textwidth]{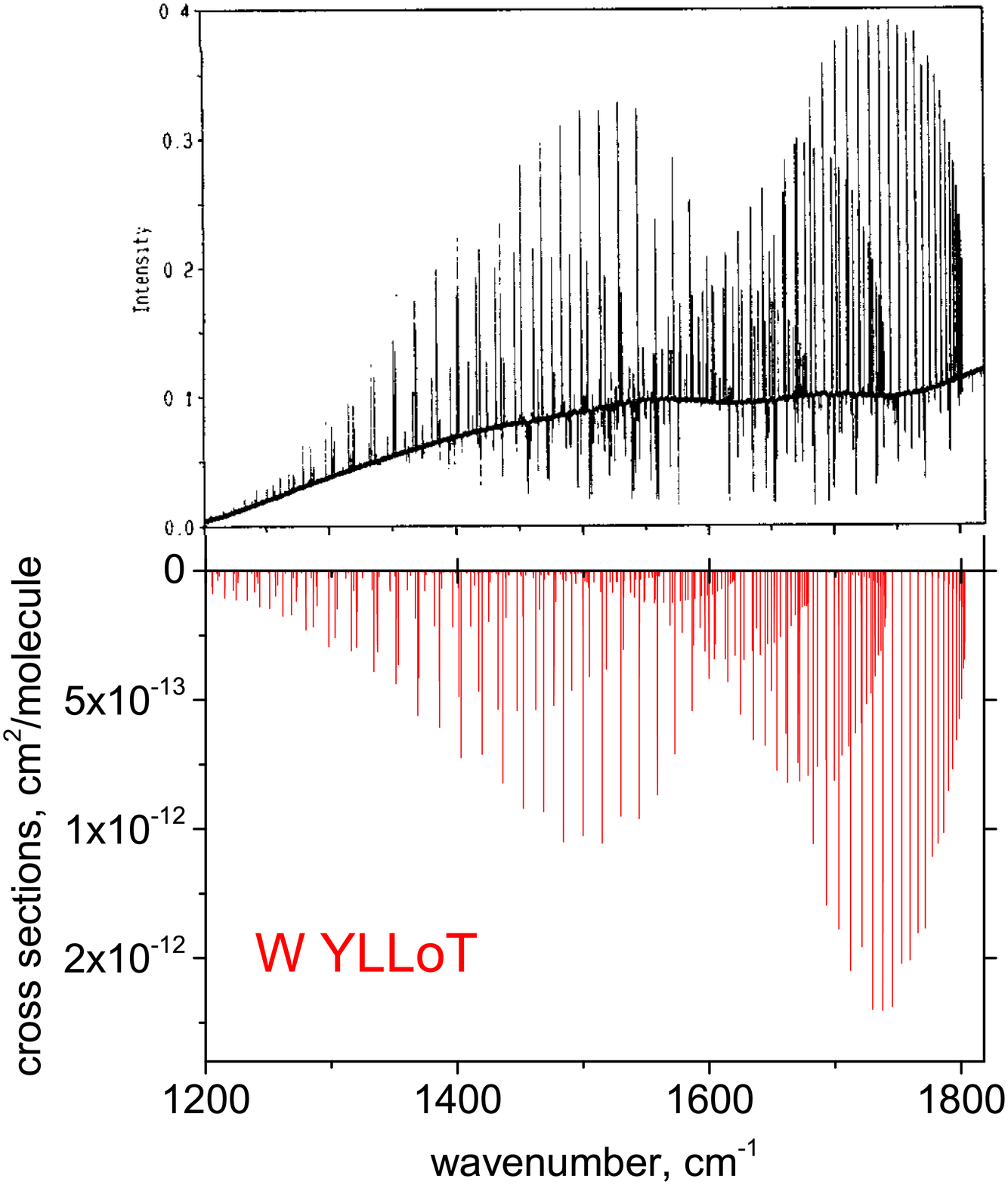}
\caption{Infrared spectrum of AlH by \citet{93WhDuBe.AlH}  (upper) compared to the emission spectrum computed using our line list assuming a temperature
of 1700~K and a Gaussian profile with the HWHM of 0.01~\cm. Reproduced from
\citet{93WhDuBe.AlH}, with the permission of AIP Publishing.}
\label{f:Bernath}
\end{center}
\end{figure}

Our line lists can be used to generate spectra for a variety of
conditions. First we compare with available laboratory spectra.
Figure~\ref{f:Bernath} compares an emission infrared spectrum of AlH
recorded by \citet{93WhDuBe.AlH} with that generated using our line
list assuming a temperature of 1700~K. Although the experimental
spectrum does not provide the absolute scale for the intensities,
there is good agreement for the relative intensities of the hot bands
in this region between the experiment and our predictions. Our
R-branch appears to be slightly stronger relative to the P-branch than
the observations of \citet{93WhDuBe.AlH}, but given the variable
baseline and presence of self-absorption in the observed spectrum this
may not be significant.

\begin{figure}
\begin{center}
\includegraphics[width=0.8\textwidth]{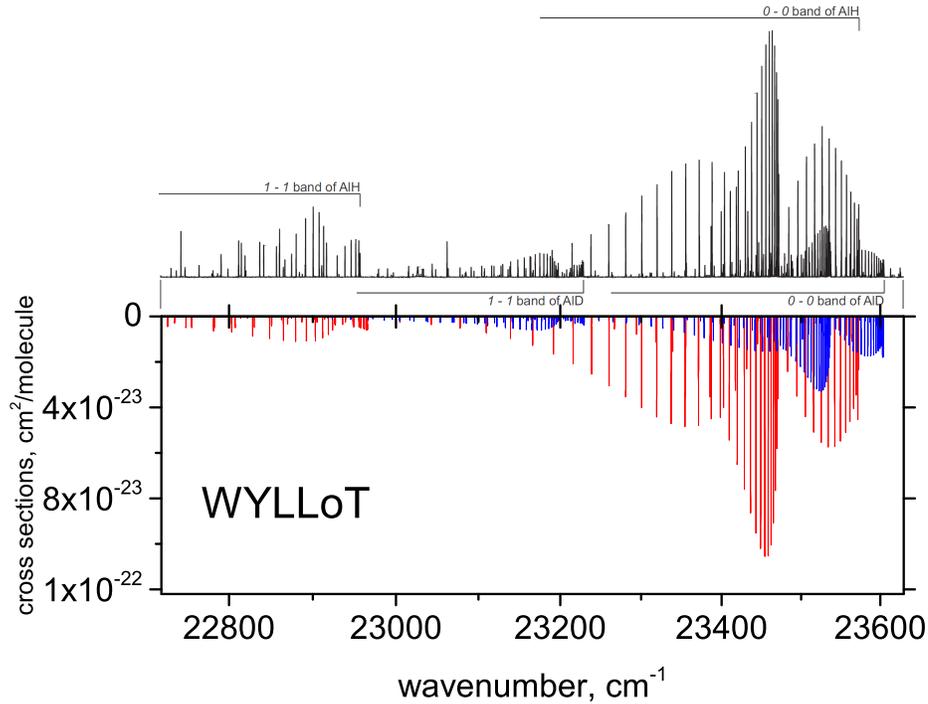}
\caption{Comparison between the 0--0 and 1--1 bands of the \A\ -- \X\
  emission spectrum of AlH and AlD measured by \citet{15SzZaHab.AlH}
  (upper) and computed using our line list (lower) assuming a
  temperature of 4500~K (vibrational) and 900~K (rotational). The
  intensities of the AlD theoretical spectrum, given in blue and
annotated in the experimental spectrum, are scaled by a
  factor 0.5.  Reprinted from \citet{15SzZaHab.AlH},
  Copyright (2015), with permission from Elsevier.}
\label{fig:15Sz}
\end{center}
\end{figure}

Figure~\ref{fig:15Sz} shows a comparison with the emission 0--0 and
1--1 bands of the \A\ -- \X\ system  of AlH and AlD by \citet{15SzZaHab.AlH}.
The observed spectrum was produced from an electric discharge in an
aluminium hollow-cathode lamp. Our spectrum has been synthesized
assuming a vibrational temperature of 4500 K and rotational
temperature of 900~K.  Comparisons with the figure suggest that the
experiments had an even lower effective rotational temperature and a
higher effective vibrational temperature. Inspection of
Fig.~\ref{fig:15Sz} suggests that our 1--1 band is blue-shifted
relative to the experiment by about 7~~\cm. Our actual numerical
agrement is much better (within experimental uncertainly), which
suggest some problems with the original figure from this paper.

\begin{figure}
\begin{center}
\includegraphics[width=0.8\textwidth]{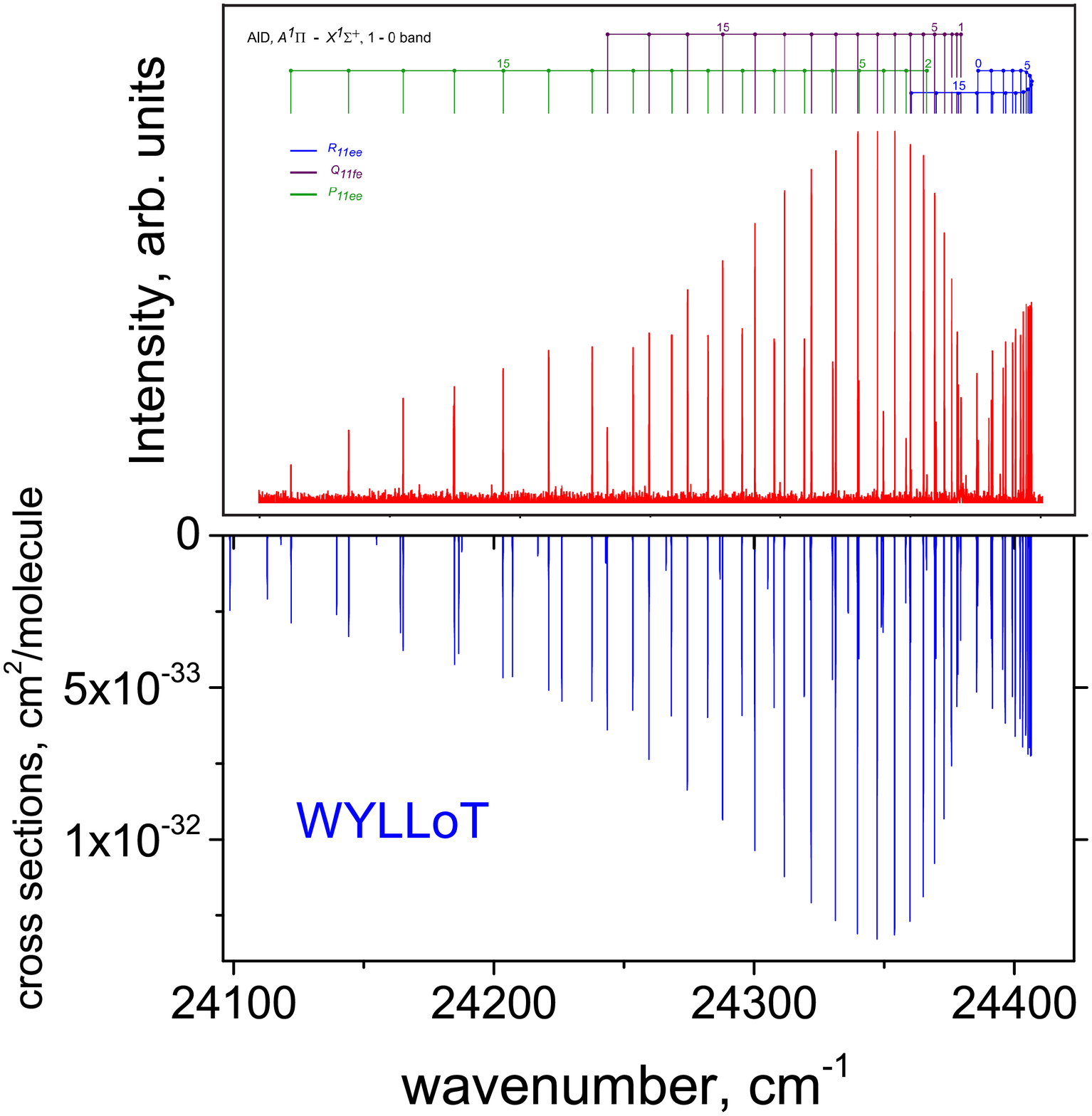}
\caption{Comparison between the 1--0 band of the \A\ -- \X\ emission spectrum of AlD
measured by \citet{17SzHaKo.AlH} (upper) and computed using our line list assuming a temperature of 600~K. Reprinted from \citet{17SzHaKo.AlH}, Copyright (2017), with permission from Elsevier.}
\label{fig:AlD:1-0}
\end{center}
\end{figure}

Figure~\ref{fig:AlD:1-0} illustrate a good agreement of the theoretical emission spectrum of the \A\ -- \X\  band $(1,0)$ at $T=600$~K of AlD with the experiment by \citet{17SzHaKo.AlH}.

\citet{92RiPaNe.AlH} reported experimentally determined ratios of Einstein coefficients for a number of vibrational bands of \A\ -- \X, which we use to assess our transition probabilities in Table~\ref{t:A}. 
Our ratios are found to be in excellent agreement with  experiment.

\begin{table}
\caption{\A\ -- \X\ Einstein coefficients $A_{v'v''} $ (s$^{-1}$) compared
to experimentally-derived ratios from \citet{92RiPaNe.AlH}, where we used $J'=1$ and $J''=0$.}
\label{t:A}
\begin{center}
\begin{tabular}{lrrrl}
\hline
\hline
                  &\multicolumn{1}{c}{This work} & \multicolumn{1}{c}{Exp.}   \\
\hline
$  A_{00}        $&$         9028000  $&$                     $\\
$  A_{01}        $&$            3878  $&$                     $\\
$  A_{10}        $&$          810100  $&$                     $\\
$  A_{12}        $&$          165900  $&$                     $\\
$  A_{11}        $&$         5407000  $&$                     $\\
$  A_{01}/A_{00}  $&$         0.00043  $&$   0.0018 \pm 0.0001 $\\
$  A_{12}/A_{11}  $&$         0.03068  $&$   0.029 \pm 0.004   $\\
$  A_{10}/A_{11}  $&$         0.14982  $&$   0.15 \pm 0.02     $\\
\hline
\end{tabular}
\end{center}
\end{table}

\begin{figure}
\begin{center}
\scalebox{0.4}{\includegraphics{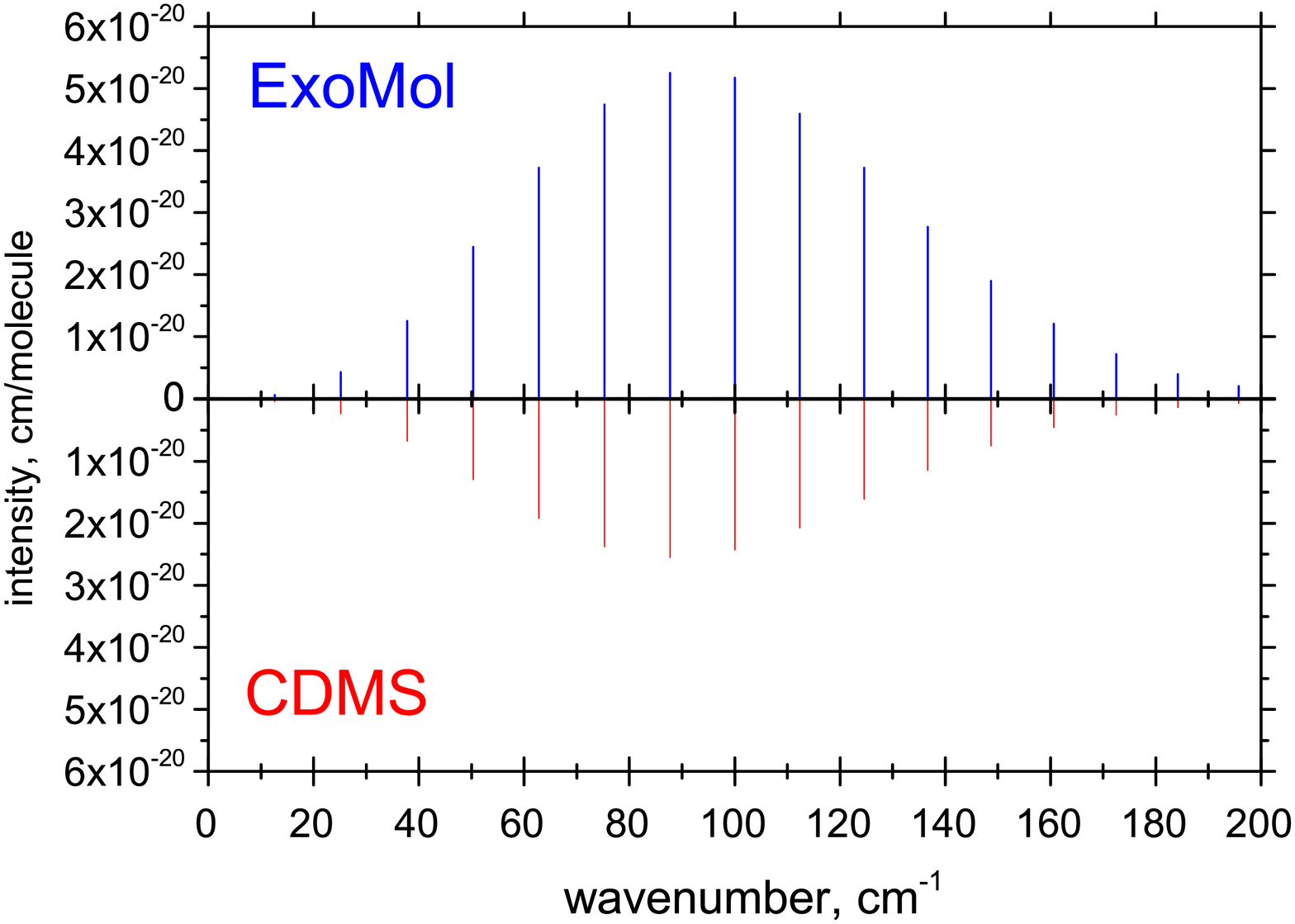}}
\caption{Comparison between pure rotational spectrum of $^{27}$AlH at 298 K given by CDMS \protect\citep{CDMS} (lower) and computed using our line list (upper).}
\label{fig:CDMS}
\end{center}
\end{figure}

Finally, Figure~\ref{fig:CDMS} gives a comparison with the
long-wavelength, rotational spectrum taken from the CDMS database
\citep{CDMS}. The agreement between the line positions is excellent,
although we recommend using the highly-accurate CDMS frequency
directly for long-wavelength studies of cool sources. However, there
is approximately a factor of two discrepancy in the predicted line
intensities. This difference is almost exactly in line with the square
of the ratio of our dipole to that of \citet{75MeRoxx.AlH} used by
CDMS. Our vibrationally averaged value for the ground state of AlH is
$0.248$~D, which is significantly different from the permanent
$\mu_{\rm e}$ value 0.158 D.  We would recommend that CDMS adopts
our value in future and note that using this value will approximately
double the upper limit for AlH in IRC+10216 determined by
\citet{10CeDeBa.HCl}.

\section{Conclusions}
\label{s:concl}

Line lists for three AlH isotopologue species  were computed using a
mixture of empirical and \ai\ curves representing our spectroscopic
model. The programs \Level\ \citep{level} and \Duo\ \citep{Duo} were
used to obtain the empirical curves, while \Duo\ was used for the line
list production. The partition functions and lifetimes were computed
using the program \xcross\ \citep{jt708}.

We provide comprehensive line lists for AlH isotopologue
species which
can be downloaded from the CDS, via
ftp://cdsarc.u-strasbg.fr/pub/cats/J/MNRAS/, or
http://cdsarc.u-strasbg.fr/viz-bin/qcat?J/MNRAS/, or from
www.exomol.com.

\section*{Acknowledgements}
We thank Nadia Milazzo and Riccardo Lanzarone for their help with the early
stages of this project.
This work was supported by the UK Science and Technology Research
Council (STFC) No. ST/M001334/1 and the COST action MOLIM No. CM1405.
This work made extensive use of UCL's Legion high performance
computing facility.

\bibliographystyle{mn2e}

\begin{thebibliography}{71}
\expandafter\ifx\csname natexlab\endcsname\relax\def\natexlab#1{#1}\fi

\bibitem[{{Allard}(2014)}]{BT-Settl}
{Allard} F., 2014, in IAU Symposium, Vol. 299, IAU Symposium, {Booth} M.,
  {Matthews} B.~C., {Graham} J.~R., eds., pp. 271--272

\bibitem[{Baltayan \& Nedelec(1979)}]{79BaNexx.AlH}
Baltayan P., Nedelec O., 1979, J. Chem. Phys., 70, 2399

\bibitem[{Barklem \& Collet(2016)}]{16BaCoxx.partfunc}
Barklem P.~S., Collet R., 2016, A\&A, 588, A96

\bibitem[{Bauschlicher \& Langhoff(1988)}]{88BaLaxx.AlH}
Bauschlicher C.~W., Langhoff S.~R., 1988, J. Chem. Phys., 89, 2116

\bibitem[{Brown \& Wasylishen({2013})}]{13BrWaxx.AlH}
Brown A., Wasylishen R.~E., {2013}, J. Mol. Spectrosc., {292}, 8

\bibitem[{Cave {et~al}\mbox{.}(1994)Cave, Johnson, \& Anderson}]{94CaJoAn.AlH}
Cave R.~J., Johnson J.~L., Anderson M.~A., 1994, Intern. J. Quantum Chem., 50,
  135

\bibitem[{Cernicharo {et~al}\mbox{.}({2010})Cernicharo, Decin, Barlow, Agundez,
  Royer, Vandenbussche, Wesson, Polehampton, De~Beck, Blommaert, Daniel,
  De~Meester, Exter, Feuchtgruber, Gear, Goicoechea, Gomez, Groenewegen,
  Hargrave, Huygen, Imhof, Ivison, Jean, Kerschbaum, Leeks, Lim, Matsuura,
  Olofsson, Posch, Regibo, Savini, Sibthorpe, Swinyard, Vandenbussche, \&
  Waelkens}]{10CeDeBa.HCl}
Cernicharo J. {et~al.}, {2010}, A\&A, {518}, L136

\bibitem[{Cobos(2002)}]{02Coxxxx.AlH}
Cobos C.~J., 2002, J. Molec. Struct. (THEOCHEM), 581, 17

\bibitem[{Deutsch {et~al}\mbox{.}(1987)Deutsch, Neil, \& Ramsay}]{87DeNeRa.AlH}
Deutsch J.~L., Neil W.~S., Ramsay D.~A., 1987, J. Mol. Spectrosc., 125, 115

\bibitem[{Diehl {et~al}\mbox{.}(2003)Diehl, Kn{\"o}dlseder, Lichti, Kretschmer,
  Schanne, Sch{\"o}nfelder, Strong, {von Kienlin}, Weidenpointner, Winkler, \&
  Wunderer}]{08DiKnLi.AlH}
Diehl R. {et~al.}, 2003, A\&A, 411, L451

\bibitem[{Furtenbacher \& {Cs\'asz\'ar}(2012)}]{12FuCsa}
Furtenbacher T., {Cs\'asz\'ar} A.~G., 2012, J. Quant. Spectrosc. Radiat.
  Transf., 113, 929

\bibitem[{Furtenbacher {et~al}\mbox{.}(2007)Furtenbacher, {Cs\'asz\'ar}, \&
  Tennyson}]{jt412}
Furtenbacher T., {Cs\'asz\'ar} A.~G., Tennyson J., 2007, J. Mol. Spectrosc.,
  245, 115

\bibitem[{Gamache {et~al}\mbox{.}(2017)Gamache, Roller, Lopes, Gordon, Rothman,
  Polyansky, Zobov, Kyuberis, Tennyson, Yurchenko, Cs\'asz\'ar, Furtenbacher,
  Huang, Schwenke, Lee, Drouin, Tashkun, I.Perevalov, \& Kochanov}]{jt692}
Gamache R.~R. {et~al.}, 2017, J. Quant. Spectrosc. Radiat. Transf., 203, 70

\bibitem[{Goto \& Saito(1995)}]{95GoSaxx.AlH}
Goto M., Saito S., 1995, ApJ, 452, L147

\bibitem[{Halfen \& Ziurys(2004)}]{04HaZixx.AlH}
Halfen D.~T., Ziurys L.~M., 2004, ApJ, 607, L63

\bibitem[{Halfen \& Ziurys(2010)}]{10HaZixx.AlH}
Halfen D.~T., Ziurys L.~M., 2010, ApJ, 713, 520

\bibitem[{Halfen \& Ziurys({2014})}]{14HaZixx.AlH}
Halfen D.~T., Ziurys L.~M., {2014}, ApJ, {791}, 65

\bibitem[{Halfen \& Ziurys({2016})}]{16HaZixx.AlH}
Halfen D.~T., Ziurys L.~M., {2016}, ApJ, {833}, 89

\bibitem[{Herbig({1956})}]{56Herbig.AlH}
Herbig G.~H., {1956}, PASP, {68}, 204

\bibitem[{Holst \& Hulth{\'e}n(1934)}]{34HoHuxx.AlH}
Holst W., Hulth{\'e}n E., 1934, Z. Phys., 90, 712

\bibitem[{Huron(1969)}]{69Huronx.AlH}
Huron B., 1969, Physica, 41, 58

\bibitem[{Ito {et~al}\mbox{.}(1994)Ito, Nakanga, Takeo, \&
  Jones}]{94ItNaTa.AlH}
Ito F., Nakanga T., Takeo H., Jones H., 1994, J. Mol. Spectrosc., 164, 379

\bibitem[{Kaminski {et~al}\mbox{.}({2016})Kaminski, Wong, Schmidt, Mueller,
  Gottlieb, Cherchneff, Menten, Keller, Bruenken, Winters, \&
  Patel}]{16KaWoSc.AlH}
Kaminski T. {et~al.}, {2016}, A\&A, {592}, A42

\bibitem[{Karthikeyan {et~al}\mbox{.}(2010)Karthikeyan, Rajamanickam, \&
  Bagare}]{10KaRaBa.AlH}
Karthikeyan B., Rajamanickam N., Bagare S.~P., 2010, Solar Phys., 264, 279

\bibitem[{{Le Roy}(2007)}]{lr07}
{Le Roy} R.~J., 2007, {LEVEL 8.0} A Computer Program for Solving the Radial
  Schr\"odinger Equation for Bound and Quasibound Levels. University of
  Waterloo Chemical Physics Research Report CP-663,
  \url{http://leroy.uwaterloo.ca/programs/}

\bibitem[{{Le Roy}(2017)}]{level}
{Le Roy} R.~J., 2017, J. Quant. Spectrosc. Radiat. Transf., 186, 167

\bibitem[{Lee {et~al}\mbox{.}(1999)Lee, Seto, Hirao, Bernath, \& Le~Roy}]{EMO}
Lee E.~G., Seto J.~Y., Hirao T., Bernath P.~F., Le~Roy R.~J., 1999, J. Mol.
  Spectrosc., 194, 197

\bibitem[{Lugaro {et~al}\mbox{.}({2012})Lugaro, Doherty, Karakas, Maddison,
  Liffman, Garcia-Hernandez, Siess, \& Lattanzio}]{12LuDoKa.AlH}
Lugaro M., Doherty C.~L., Karakas A.~I., Maddison S.~T., Liffman K.,
  Garcia-Hernandez D.~A., Siess L., Lattanzio J.~C., {2012}, Meteorics Planet.
  Sci., {47}, 1998

\bibitem[{{Mahoney} {et~al}\mbox{.}(1984){Mahoney}, {Ling}, {Wheaton}, \&
  {Jacobson}}]{84MaLiWh.AlH}
{Mahoney} W.~A., {Ling} J.~C., {Wheaton} W.~A., {Jacobson} A.~S., 1984, ApJ,
  286, 578

\bibitem[{Matos {et~al}\mbox{.}(1988)Matos, Roos, Sadlej, \&
  Diercksen}]{88MaRoSa.AlH}
Matos J. M.~O., Roos B.~O., Sadlej A.~J., Diercksen G. H.~F., 1988, Chem.
  Phys., 119, 71

\bibitem[{Medvedev {et~al}\mbox{.}(2016)Medvedev, Meshkov, Stolyarov, Ushakov,
  \& Gordon}]{16MeMeSt}
Medvedev E.~S., Meshkov V.~V., Stolyarov A.~V., Ushakov V.~G., Gordon I.~E.,
  2016, J. Mol. Spectrosc., 330, 36

\bibitem[{Meyer \& Rosmus(1975)}]{75MeRoxx.AlH}
Meyer W., Rosmus P., 1975, J. Chem. Phys., 63, 2356

\bibitem[{{M\"uller} {et~al}\mbox{.}(2005){M\"uller}, {Schl\"oder}, Stutzki, \&
  Winnewisser}]{CDMS}
{M\"uller} H. S.~P., {Schl\"oder} F., Stutzki J., Winnewisser G., 2005, J.
  Molec. Struct. (THEOCHEM), 742, 215

\bibitem[{Nizamov \& Dagdigian(2000)}]{00NiDaxx.AlH}
Nizamov B., Dagdigian P.~J., 2000, J. Chem. Phys., 113, 4124

\bibitem[{Patrascu {et~al}\mbox{.}(2014)Patrascu, Hill, Tennyson, \&
  Yurchenko}]{jt589}
Patrascu A.~T., Hill C., Tennyson J., Yurchenko S.~N., 2014, J. Chem. Phys.,
  141, 144312

\bibitem[{Patrascu {et~al}\mbox{.}(2015)Patrascu, Tennyson, \&
  Yurchenko}]{jt598}
Patrascu A.~T., Tennyson J., Yurchenko S.~N., 2015, MNRAS, 449, 3613

\bibitem[{Prajapat {et~al}\mbox{.}(2017)Prajapat, Jagoda, Lodi, Gorman,
  Yurchenko, \& Tennyson}]{jt703}
Prajapat L., Jagoda P., Lodi L., Gorman M.~N., Yurchenko S.~N., Tennyson J.,
  2017, MNRAS, 472, 3648

\bibitem[{Rafi {et~al}\mbox{.}(1978)Rafi, Baig, \& Khan}]{78RaBaKh.AlH}
Rafi M., Baig M.~A., Khan M.~A., 1978, Nouvo Cimento Soc. Ital. Fis. B-Gen.
  Phys. Relativ. Astron. Math. Phys. Methods, 43, 271

\bibitem[{Rajpurohit {et~al}\mbox{.}({2013})Rajpurohit, Reyle, Allard, Homeier,
  Schultheis, Bessell, \& Robin}]{13RaReAl.NaHAlH}
Rajpurohit A.~S., Reyle C., Allard F., Homeier D., Schultheis M., Bessell
  M.~S., Robin A.~C., {2013}, A\&A, {556}, A15

\bibitem[{Ram \& Bernath(1996)}]{96RaBexx.AlH}
Ram R.~S., Bernath P.~F., 1996, Appl. Optics, 35, 2879

\bibitem[{Rice {et~al}\mbox{.}(1992)Rice, Pasternack, \& Nelson}]{92RiPaNe.AlH}
Rice J.~K., Pasternack L., Nelson H.~H., 1992, Chem. Phys. Lett., 189, 43

\bibitem[{Rivlin {et~al}\mbox{.}(2015)Rivlin, Lodi, Yurchenko, Tennyson, \& {Le
  Roy}}]{jt605}
Rivlin T., Lodi L., Yurchenko S.~N., Tennyson J., {Le Roy} R.~J., 2015, MNRAS,
  451, 5153

\bibitem[{Sauval \& Tatum(1984)}]{84SaTaxx.partfunc}
Sauval A.~J., Tatum J.~B., 1984, ApJS, 56, 193

\bibitem[{Seck {et~al}\mbox{.}(2014)Seck, Hohenstein, Lien, Stollenwerk, \&
  Odom}]{14SeHoLi.AlH}
Seck C.~M., Hohenstein E.~G., Lien C.-Y., Stollenwerk P.~R., Odom B.~C., 2014,
  J. Mol. Spectrosc., 300, 108

\bibitem[{Szajna {et~al}\mbox{.}(2017{\natexlab{a}})Szajna, Hakalla, Kolek, \&
  Zachwieja}]{17SzHaKo.AlH}
Szajna W., Hakalla R., Kolek P., Zachwieja M., 2017{\natexlab{a}}, J. Quant. Spectrosc.
  Radiat. Transf., {187}, 167

\bibitem[{Szajna {et~al}\mbox{.}(2017{\natexlab{b}})Szajna, Moore, \& Lane}]{17SzMoLa.AlH}
Szajna W., Moore K., Lane I.~C., 2017{\natexlab{b}}, J. Quant. Spectrosc. Radiat. Transf.,
  196, 103

\bibitem[{Szajna \& Zachwieja(2009)}]{09SzZaxx.AlH}
Szajna W., Zachwieja M., 2009, Eur. Phys. J. D, 55, 549

\bibitem[{Szajna \& Zachwieja(2010)}]{10SzZaxx.AlH}
Szajna W., Zachwieja M., 2010, J. Mol. Spectrosc., 260, 130

\bibitem[{Szajna \& Zachwieja(2011)}]{11SzZaxx.AlH}
Szajna W., Zachwieja M., 2011, J. Mol. Spectrosc., 269, 56

\bibitem[{Szajna {et~al}\mbox{.}({2015})Szajna, Zachwieja, \&
  Hakalla}]{15SzZaHab.AlH}
Szajna W., Zachwieja M., Hakalla R., {2015}, J. Mol. Spectrosc., {318}, 78

\bibitem[{Szajna {et~al}\mbox{.}(2011)Szajna, Zachwieja, Hakalla, \&
  Kepa}]{11SzZaHa.AlH}
Szajna W., Zachwieja M., Hakalla R., Kepa R., 2011, Acta Phys. Pol.A, 120, 417

\bibitem[{Tao {et~al}\mbox{.}(2003)Tao, Tan, Dagdigian, \&
  Alexander}]{03TaTaDa.AlH}
Tao C., Tan X.~F., Dagdigian P.~J., Alexander M.~H., 2003, J. Chem. Phys., 118,
  10477

\bibitem[{Tennyson(2014)}]{jt573}
Tennyson J., 2014, J. Mol. Spectrosc., 298, 1

\bibitem[{Tennyson {et~al}\mbox{.}(2016{\natexlab{a}})Tennyson, Lodi,
  McKemmish, \& Yurchenko}]{jt632}
Tennyson J., Lodi L., McKemmish L.~K., Yurchenko S.~N., 2016{\natexlab{a}}, J.
  Phys. B: At. Mol. Opt. Phys., 49, 102001

\bibitem[{Tennyson \& Yurchenko(2012)}]{jt528}
Tennyson J., Yurchenko S.~N., 2012, MNRAS, 425, 21

\bibitem[{Tennyson \& Yurchenko(2017)}]{jt693}
Tennyson J., Yurchenko S.~N., 2017, Mol. Astrophys., 8, 1

\bibitem[{Tennyson {et~al}\mbox{.}(2016{\natexlab{b}})Tennyson, Yurchenko,
  Al-Refaie, Barton, Chubb, Coles, Diamantopoulou, Gorman, Hill, Lam, Lodi,
  McKemmish, Na, Owens, Polyansky, Rivlin, Sousa-Silva, Underwood, Yachmenev,
  \& Zak}]{jt631}
Tennyson J. {et~al.}, 2016{\natexlab{b}}, J. Mol. Spectrosc., 327, 73

\bibitem[{Urban \& Jones(1992)}]{92UrJoxx.AlH}
Urban R.~D., Jones H., 1992, Chem. Phys. Lett., 190, 609

\bibitem[{Vidler \& Tennyson(2000)}]{jt263}
Vidler M., Tennyson J., 2000, J. Chem. Phys., 113, 9766

\bibitem[{\v{S}urkus {et~al}\mbox{.}(1984)\v{S}urkus, Rakauskas, \&
  Bolotin}]{84SuRaBo.method}
\v{S}urkus A.~A., Rakauskas R.~J., Bolotin A.~B., 1984, Chem. Phys. Lett., 105,
  291

\bibitem[{Wallace {et~al}\mbox{.}(2000)Wallace, Hinkle, \&
  Livingston}]{00WaHiLi}
Wallace L., Hinkle K., Livingston W., 2000, {An Atlas of Sunspot Umbral Spectra
  in the Visible from 15 000 to 25 000 cm$^{-1}$ (3920 to 6664 \AA)}. Tech.
  Rep. Tech. Rep. 00-001, National Solar Observatory, Tucson, AZ

\bibitem[{Wells \& Lane(2011)}]{11WeLaxx.AlH}
Wells N., Lane I.~C., 2011, Phys. Chem. Chem. Phys., 13, 19018

\bibitem[{Werner {et~al}\mbox{.}(2012)Werner, Knowles, Knizia, Manby, \&
  Sch\"utz}]{MOLPRO}
Werner H.-J., Knowles P.~J., Knizia G., Manby F.~R., Sch\"utz M., 2012, WIREs
  Comput. Mol. Sci., 2, 242

\bibitem[{White {et~al}\mbox{.}(1993)White, Dulick, \& Bernath}]{93WhDuBe.AlH}
White J.~B., Dulick M., Bernath P.~F., 1993, J. Chem. Phys., 99, 8371

\bibitem[{Yamada \& Hirota(1992)}]{92YaHixx.AlH}
Yamada C., Hirota E., 1992, Chem. Phys. Lett., 197, 461

\bibitem[{Yang \& Dagdigian(1998)}]{98YaDaxx.AlH}
Yang X., Dagdigian P.~J., 1998, J. Chem. Phys., 109, 8920

\bibitem[{Yurchenko {et~al}\mbox{.}(2018{\natexlab{a}})Yurchenko, Al-Refaie, \&
  Tennyson}]{jt708}
Yurchenko S.~N., Al-Refaie A.~F., Tennyson J., 2018{\natexlab{a}}, A\&A, (in
  press)

\bibitem[{Yurchenko {et~al}\mbox{.}(2016)Yurchenko, Lodi, Tennyson, \&
  Stolyarov}]{Duo}
Yurchenko S.~N., Lodi L., Tennyson J., Stolyarov A.~V., 2016, Comput. Phys.
  Commun., 202, 262

\bibitem[{Yurchenko {et~al}\mbox{.}(2018{\natexlab{b}})Yurchenko, Sinden, Lodi,
  Hill, Gorman, \& Tennyson}]{jt711}
Yurchenko S.~N., Sinden F., Lodi L., Hill C., Gorman M.~N., Tennyson J.,
  2018{\natexlab{b}}, MNRAS, 473, 5324

\bibitem[{Zhang \& Stuke(1988)}]{88ZhStxx.AlH}
Zhang Y., Stuke M., 1988, Chem. Phys. Lett., 149, 310

\bibitem[{Zhu {et~al}\mbox{.}(1992)Zhu, Shehadeh, \& Grant}]{92ZhShGr.AlH}
Zhu Y.~F., Shehadeh R., Grant E.~R., 1992, J. Chem. Phys., 97, 883

\end{thebibliography}

\label{lastpage}

\end{document}